\documentclass[10pt,conference]{IEEEtran}
\IEEEoverridecommandlockouts

\usepackage{amsmath,amssymb}
\usepackage{bm}
\usepackage{graphicx}
\usepackage{ascmac}
\usepackage{float}
\usepackage{url}
\usepackage{comment}
\usepackage{tabularx,theorem}
\usepackage{indentfirst}
\usepackage{algorithmic}
\usepackage{algorithm}
\usepackage{enumerate}
\usepackage{color}
\usepackage{physics}
\usepackage{mathtools}
\usepackage{qcircuit}
\usepackage{svg}

\usepackage[colorlinks=true,citecolor=blue,linkcolor=magenta]{hyperref}
\usepackage{cite}
\usepackage{textcomp}
\usepackage{xcolor}
\def\BibTeX{{\rm B\kern-.05em{\sc i\kern-.025em b}\kern-.08em
    T\kern-.1667em\lower.7ex\hbox{E}\kern-.125emX}}

\newcommand{\update}[1]{#1}
\begin{document}

\title{Locality-aware Pauli-based computation for local magic state preparation
\thanks{\copyright 2025 IEEE.  Personal use of this material is permitted.  Permission from IEEE must be obtained for all other uses, in any current or future media, including reprinting/republishing this material for advertising or promotional purposes, creating new collective works, for resale or redistribution to servers or lists, or reuse of any copyrighted component of this work in other works.}
}

\author{
\IEEEauthorblockN{Yutaka Hirano\IEEEauthorrefmark{1} and Keisuke Fujii\IEEEauthorrefmark{1}\IEEEauthorrefmark{2}\IEEEauthorrefmark{3}\IEEEauthorrefmark{4} 
}

\IEEEauthorblockA{
\IEEEauthorrefmark{1}\textit{Graduate School of Engineering Science}, \textit{Osaka University},\\
1-3 Machikaneyama, Toyonaka, Osaka 560-8531, Japan}

\IEEEauthorblockA{
\IEEEauthorrefmark{2}\textit{School of Engineering Science}, \textit{Osaka University},\\
1-3 Machikaneyama, Toyonaka, Osaka 560-8531, Japan}

\IEEEauthorblockA{
\IEEEauthorrefmark{3}\textit{Center for Quantum Information and Quantum Biology}, \textit{Osaka University},\\
1-2 Machikaneyama, Toyonaka 560-0043, Japan}

\IEEEauthorblockA{
\IEEEauthorrefmark{4}\textit{RIKEN Center for Quantum Computing (RQC)},
Hirosawa 2-1, Wako, Saitama 351-0198, Japan}

u965281c@ecs.osaka-u.ac.jp, fujii@qc.ee.es.osaka-u.ac.jp
}

\maketitle

\begin{abstract}
Magic state distillation, a process for preparing magic states needed to implement non-Clifford gates fault-tolerantly, plays a crucial role in fault-tolerant quantum computation.
Historically, it has been a major bottleneck, leading to the pursuit of computation schemes optimized for slow magic state preparation.
Recent advances in magic state distillation have significantly reduced the overhead, enabling the simultaneous preparation of many magic states.
However, the magic state transfer cost prevents the conventional layout from efficiently utilizing them, highlighting the need for an alternative scheme optimized for highly parallel quantum algorithms.
In this study, we propose locality-aware Pauli-based computation, a novel compilation scheme that distills magic states in the computation area, aiming to reduce execution time by minimizing magic state transfer costs and improving locality.
Numerical experiments on random circuit sampling and 2D Ising Hamiltonian simulation demonstrate that our scheme significantly reduces execution time—while incurring little or no additional spatial overhead—compared to sequential Pauli-based computation, a conventional computation scheme, and scales favorably with increasing qubit count.
\end{abstract}


\maketitle

\section{Introduction}
\label{sec:introduction}
Quantum computers promise to solve problems of practical importance that are believed to be intractable for classical computers—for example, simulating quantum physics~\cite{Abrams1999Quantum} and integer factoring~\cite{Shor1994Factoring}.
Existing quantum computers, often referred to as noisy intermediate-scale quantum (NISQ) devices, are highly susceptible to hardware errors, preventing them from executing complex quantum algorithms such as Shor's factoring algorithm~\cite{Shor1994Factoring}.
To fully realize this promise, we need fault-tolerant quantum computers (FTQC) based on quantum error correction (QEC), which protects information from errors by encoding \textit{logical qubits} into many physical qubits.
Surface codes~\cite{Kitaev2003,Bravyi1998} are among the most promising candidates for QEC because of their relatively high error threshold and compatibility with nearest-neighbor connectivity.
Experimental quantum error correction has been demonstrated using the surface code architecture in quantum devices with nearest-neighbor interactions, such as superconducting qubits~\cite{Arya2024QuantumErrorCorrection}.

To perform computations on an FTQC, we need \textit{logical gates} applied to logical qubits.
For the surface code, the Clifford+$T$ gate set, consisting of $\{H, S, T, CX\}$ gates, is often adopted to enable universal quantum computation.
Among these, the Clifford gates $\{H, S, CX\}$ are relatively inexpensive, whereas the $T$ gate is more resource-intensive.
To perform a $T$ gate, we prepare a \textit{magic state}, $T\ket{+}$, through a process called \textit{magic state distillation}~\cite{Bravyi2005}.
Magic state distillation is expensive and has long been considered a major bottleneck in fault-tolerant quantum computation.
This has led to the pursuit of compilation schemes optimized for the high cost of magic state preparation, where a compilation scheme consists of the \textit{transpilation} process, \textit{layout}, and other compilation algorithms. 

\begin{figure}[t!]
\centering
\includesvg[width=6cm]{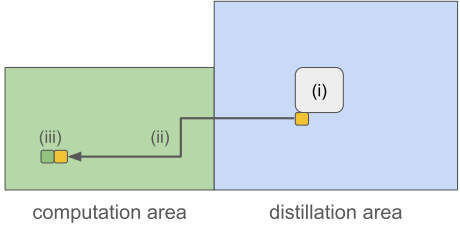}
\caption{Performing a T gate in the conventional layout.
(i) Magic state distillation.
(ii) Magic state transfer from the distillation area (blue) to the computation area (green).
(iii) Gate teleportation.}
\label{fig:conventional-layout}
\end{figure}

Sequential Pauli-based computation (SPC)~\cite{Litinski2019GameOfSurfaceCodes} is one such scheme.
It transpiles a sequence of Clifford+$T$ gates into multi-qubit Pauli measurements.
SPC uses a layout that separates the computation area from the distillation area.
In this layout, executing a $T$ gate involves three steps: (i) distilling a magic state in the distillation area, (ii) transferring it to a qubit adjacent to the target data qubit in the computation area, and (iii) performing gate teleportation (\autoref{fig:conventional-layout}).
This layout has been widely adopted in resource estimation and compilation studies for FTQC with nearest-neighbor connectivity~\cite{Beverland2022EdgeDisjointPath, Beverland2022Assessing, Chamberland2022Universal, leblond2024realistic, Molavi2024Dependency, Hamada2024Efficient}.

Recent advances have significantly reduced the overhead of magic state distillation~\cite{Goto2016Minimizing, Litinski2019magicstate, Itogawa2024Efficient, Hirano2024Leveraging, Smith2024Mitigating, Gidney2024MagicStateCultivation}, enabling the simultaneous preparation of many magic states.
However, the cost of magic state transfer—which scales with the length of the transfer path—prevents the conventional layout from efficiently utilizing these states.
This highlights the need for an alternative layout and a transpilation strategy optimized for highly parallel quantum algorithms.

In this work, we propose locality-aware Pauli-based computation, a novel FTQC compilation scheme optimized for highly parallel quantum algorithms.
More specifically, we propose a compilation scheme that performs magic state distillation locally in the computational area, which has recently been greatly optimized with respect to the spatial cost.
To make full use of this locality, our scheme transpiles an input quantum algorithm composed of Clifford+$T$ gates into a sequence of single-qubit and two-qubit Pauli measurements with magic states, while preserving the gate locality of the input algorithm.
The combination of gate locality and local magic state preparation enables the parallel execution of many $T$ gates.
High-performance distillation techniques rely on post-selection to reduce overhead, resulting in low acceptance rates.
We mitigate this by running multiple parallel distillation processes for each magic state.
The enhanced locality offers greater parallelism than existing parallel-computing schemes based on the conventional layout~\cite{Beverland2022EdgeDisjointPath, Beverland2022Assessing, Molavi2024Dependency}.
Moreover, our scheme is compatible with many routing and scheduling algorithms developed for conventional layouts, opening the door to even more efficient compilation strategies.

To evaluate the scheme's performance, we developed a scheduler and a runtime simulator that implement locality-aware Pauli-based computation.
We performed numerical simulations of two algorithms—random circuit sampling and 2D Ising Hamiltonian simulation—using the scheduler and simulator.
For 6$\times$6 (12$\times$12) random quantum circuit sampling, we observed a 14\% (71\%) reduction in execution time compared to SPC, with little or no additional spatial overhead.
For 6$\times$6 (12$\times$12) Hamiltonian simulations, execution time was reduced by 48\% (84\%) compared to SPC, again with little or no additional spatial overhead.
These results demonstrate that locality-aware Pauli-based computation scales favorably in execution time as the number of qubits increases.
Therefore, we conclude that locality-aware Pauli-based computation will become a standard compilation scheme targeting highly parallel quantum algorithms when high-performance magic state distillation is available.
Our scheduler and runtime simulator reveal the relationship between the magic state distillation cost---including the acceptance ratio---and the overall computational cost for a given quantum algorithm.  
This provides useful design guidelines for developing magic state distillation protocols.

The rest of this paper is organized as follows.
In \autoref{sec:preliminary}, we provide basic definitions, notations, and existing studies essential for understanding our proposal.
In \autoref{sec:proposal}, we detail our proposal, locality-aware Pauli-based computation, including theoretical analyses and comparisons with other schemes.
In \autoref{sec:performance-evaluation}, we conduct a performance evaluation through resource estimation with random circuit sampling and Hamiltonian simulation with Trotterization.
Finally, \autoref{sec:conclusion} concludes the paper with a summary of our findings.

\section{Preliminary}
\label{sec:preliminary}
\subsection{Basic operators}
\label{subsec:preliminary-basic-operators}
We begin by defining important classes of operators, as they play a crucial role in the following discussions.
\textit{Pauli operators} are defined as the one-qubit Pauli operators ($I$, $X$, $Y$, and $Z$) and their tensor products, together with factors $\pm1$ and ${\pm}i$.
Pauli operators acting on $n$ qubits, denoted by $\mathcal{P}_n$, form a group.
Clifford operators are defined as unitary operators that map a Pauli operator to a Pauli operator through conjugation.
\[
\mathcal{C}_n \coloneqq \{U \mid U\mathcal{P}_nU^{\dagger} \subset \mathcal{P}_n \}.
\]
This implies that a Pauli gate $P$ followed by a Clifford gate $C$ is equivalent to $C$ followed by another Pauli gate, as illustrated in \autoref{fig:pauli-feedforward}.
By leveraging this property, we can move all Pauli gates in a Clifford circuit to the end without modifying the non-Pauli portion of the circuit.
This process, known as \textit{Pauli feedforwarding}, can be efficiently simulated by classical computers~\cite{riesebos2017pauli}.
In this study, we make extensive use of Pauli feedforwarding and may ignore Pauli gates in subsequent discussions.

\begin{figure}[t!]
\centering
 \[
 \Qcircuit @C=.4em @R=.8em {
   \lstick{} & \qw & \gate{P} & \gate{C} & \qw & \push{\rule{0.6em}{0em}=\rule{0.6em}{0em}} & \qw & \gate{C} & \gate{C^\dagger} & \gate{P} & \gate{C} & \qw & \push{\rule{0.6em}{0em}=\rule{0.6em}{0em}} & \qw & \gate{C} & \gate{CPC^{\dagger}} & \qw \\
   }
 \]
\caption{
A Pauli gate $P$ followed by a Clifford gate $C$ is equivalent to $C$ followed by another Pauli gate.
}
\label{fig:pauli-feedforward}
\end{figure}

We define a Pauli rotation along a Pauli operator $P$ with a rotation angle $\theta$,
\[
e^{i \theta P} = \cos \theta I + i \sin \theta P.
\]
$e^{i \theta P}$ is a Pauli operator if and only if $\theta$ is a multiple of $\frac{\pi}{2}$, and a Clifford operator if and only if $\theta$ is a multiple of $\frac{\pi}{4}$.
We refer to $e^{i \theta P}$ as a $\frac{\pi}{8}$ or $\frac{\pi}{4}$ rotation if $\theta \in \{\frac{\pi}{8}, -\frac{\pi}{8}\}$ or $\{\frac{\pi}{4}, -\frac{\pi}{4}\}$, respectively.
We refer to $M_P = I + (-1)^{m}P$ as a \textit{Pauli measurement}, where $P$ is a Pauli operator and $m \in \{0, 1\}$ is the measurement outcome.

\subsection{The surface code}
\label{subsec:preliminary-surface-code}
In this study, we focus on quantum computers with nearest-neighbor connectivity, such as superconducting quantum computers.
We use the rotated surface code~\cite{Kitaev2003,Bravyi1998}, which has a relatively high error threshold and supports nearest-neighbor connectivity.

\begin{figure}[t!]
\centering
\includesvg[width=1.5cm]{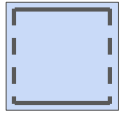}
\caption{
A surface code patch.
Solid and dashed lines represent logical $Z$ and $X$ operators, respectively.
These lines may be omitted when not relevant to computation.}
\label{fig:surface-code-path}
\end{figure}

With the surface code, multiple physical qubits form one logical qubit.
A set of physical qubits encoding a logical qubit is called a \textit{surface code patch}.
A surface code patch has two edges corresponding to its logical $Z$ operator and two edges corresponding to its logical $X$ operator, depicted as solid and dashed lines in figures, respectively (\autoref{fig:surface-code-path}).
In our model, a quantum computer consists of surface code patches placed on a 2D grid.
We use the time required to perform an error syndrome measurement (\textit{cycle time}) as the time unit.
For example, performing a lattice surgery operation~\cite{Horsman2012} requires $d$ cycles, where $d$ is the code distance.

\subsection{FTQC compilation}
\label{subsec:preliminary-ftqc-compiler}
A fault-tolerant quantum computer is a complex system consisting of many layers.
In this study, we model FTQC as a sequence of layers, ordered from high level to low level.
This hierarchical model is based on the model proposed in~\cite{Beverland2022Assessing}.

\begin{enumerate}
  \item \textbf{Quantum high-level programming language}: A high-level programming language such as Q\#~\cite{Svore2018QSharp} and Quipper~\cite{Green2013Quipper} for describing quantum algorithms.
  \item \textbf{Quantum IR}: An intermediate representation (IR) such as QIR~\cite{QIRSpec2021} that lacks high-level language features.
  \item \textbf{Quantum ISA}: A set of fundamental operations (e.g., lattice surgery~\cite{Horsman2012}) that implement quantum IR instructions.
  \item \textbf{QEC code}:  As discussed in \autoref{subsec:preliminary-surface-code}, we use the rotated surface code throughout this study.
  \item \textbf{Hardware}: A quantum device accompanied by classical computers on which a QEC code is implemented.
\end{enumerate}

In general, any translation process from an upper layer to a lower layer can be called compilation.
However, in this study, we use the term \textit{compilation} specifically for a translation process from the quantum IR layer to the QEC code layer.
We also refer to \textit{transpilation} as a translation process from a quantum IR to a quantum ISA.
We define a \textit{compilation scheme} as a quantum ISA specification accompanied by a compilation algorithm.
Sequential Pauli-based computation (\autoref{subsec:preliminary-spc}) is an example of a compilation scheme.
In this study, we assume that a quantum IR consists of the following operations:
\begin{enumerate}
 \item Single-qubit $\ket{0}$ initialization,
 \item Single-qubit $Z$ measurement,
 \item $H, S$, and $CX$ Clifford gates,
 \item $e^{i \theta Z}$, where $\theta \in [0, 2\pi)$, and
 \item Classical control.
\end{enumerate}
Although classical control is an important component of quantum computation, we omit it in this study to simplify the discussion.

\begin{figure}[t!]
\centering
\includesvg[width=5cm]{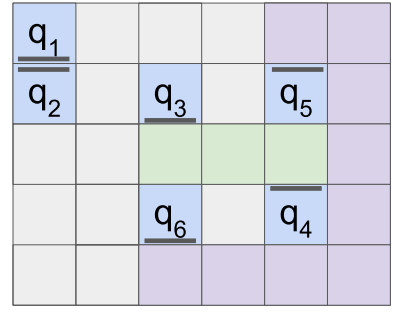}
\caption{
An example of routing.
Lattice surgery can be performed directly on $q_1$ and $q_2$ because they are adjacent and their logical $Z$ operators are aligned.
The green region represents a path required to perform $M_{Z_3Z_4}$.
The purple region depicts a path required to perform $M_{Z_5Z_6}$ without conflicting with the green path.}
\label{fig:routing}
\end{figure}

Among the various FTQC compilation features, we focus on four: \textit{routing}, \textit{mapping}, \textit{scheduling}, and \textit{gate synthesis}.
Because our target quantum computer has nearest-neighbor connectivity, performing a two-qubit logical gate requires a path between the patches corresponding to the two logical qubits.
Suppose we wish to perform a $ZZ$ measurement on two logical qubits.
If the two qubits are adjacent, we can perform lattice surgery (\autoref{fig:routing}, $q_1$ and $q_2$).
Otherwise, an intermediate path of patches must be created between them to perform a $ZZ$ measurement (\autoref{fig:routing}, $q_3$ and $q_4$).
There may be multiple possible paths between a pair of qubits, and the compiler selects one.
This process is referred to as \textit{routing}.

The computational cost of performing a two-qubit gate depends on the positions of the target qubits.
Thus, the assignment of a surface code patch to a logical qubit affects the total execution time.
We refer to this association process as \textit{mapping}.

\begin{figure}[t!]
\centering
\includesvg[width=4cm]{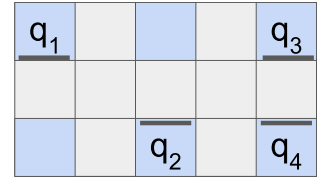}
\caption{
An example of scheduling.
Because $CZ_{14}$ and $CZ_{23}$ cannot be executed in parallel, the compiler determines their execution order.}
\label{fig:scheduling}
\end{figure}

The compiler also assigns a time slot to each operation, ensuring that the semantics of the input program are preserved.
For example, in \autoref{fig:scheduling}, $M_{Z_1Z_4}$ and $M_{Z_2Z_3}$ cannot be performed simultaneously because $M_{ZZ}$ requires a path that connects the logical $Z$ operators of the target qubits, and no two paths can simultaneously connect $q_1$ to $q_4$ and $q_2$ to $q_3$ without intersecting.
Since $M_{Z_1Z_4}$ and $M_{Z_2Z_3}$ commute, the compiler can determine their execution order.
We call this process \textit{scheduling}.

Some algorithms require arbitrary-angle $Z$ rotations of the form $e^{i\theta Z}$, where $\theta \in [0, 2\pi)$.
\textit{Gate synthesis} is the process of approximating an arbitrary-angle rotation using a circuit consisting of Clifford, $T$, initialization, and measurement gates.
In this study, we use the mixed diagonal algorithm~\cite{Kliuchnikov2023Shorter}, which approximates an arbitrary-angle rotation by decomposing it into a sequence of Clifford and $T$ gates.
The sequence contains approximately $1.5\delta$ $T$ gates, where $2^{-\delta}$ denotes the precision of the rotation angle.
Although other algorithms, such as the mixed fallback algorithm~\cite{Kliuchnikov2023Shorter}, achieve a lower $T$ count, we use the mixed diagonal algorithm for its simpler circuit structure.
We believe that the overall conclusion of this study holds regardless of the choice of a gate synthesis algorithm.

\subsection{Sequential Pauli-based computation}
\label{subsec:preliminary-spc}
In this subsection, we describe sequential Pauli-based computation, a compilation scheme introduced by Litinski~\cite{Litinski2019GameOfSurfaceCodes}.
We first define the quantum ISA and its implementation, then describe the algorithm for translating the quantum IR to the quantum ISA.

\subsubsection{Quantum ISA and its implementation}
\label{subsec:preliminary-spc-isa}
The quantum ISA for sequential Pauli-based computation consists of the following operations:
\begin{enumerate}
 \item Single-qubit $\ket{0}$ initialization,
 \item Single-qubit Pauli measurement,
 \item Multi-qubit Pauli measurement, and
 \item $\frac{\pi}{8}$ rotation along a Pauli operator $P$.
\end{enumerate}

\begin{figure}[t!]
\centering
\includesvg[width=3cm]{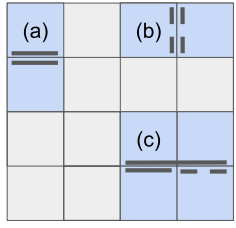}
\caption{
Multi-qubit Pauli measurement examples with lattice surgery depicting (a) $M_{ZZ}$, (b) $M_{XX}$, and (c) $M_{ZY}$.}
\label{fig:lattice-surgery}
\end{figure}

\begin{figure}[t!]
\centering
\includesvg[width=4cm]{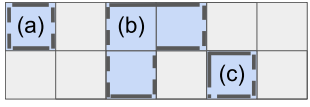}
\caption{
Qubit initialization examples. (a) Single-patch initialization. (b) Multi-patch and multi-qubit initialization. (c) Single-patch initialization with a shortened $X$ boundary. }
\label{fig:qubit-initialization}
\end{figure}

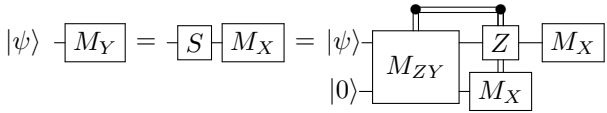
\begin{figure}[t!]
\centering
\[
\Qcircuit @C=.4em @R=.4em {
\lstick{ }          &              &          & &          &              &          &                   & \control[1] \cwx[1]   & \cctrl{1} \cwx[1]   &             \\
\lstick{\ket{\psi}} & \gate{M_{Y}} & \push{=} & & \gate{S} & \gate{M_{X}} & \push{=} & \push{\ket{\psi}} & \multigate{1}{M_{ZY}} & \gate{Z}            & \gate{M_X}  \\
\lstick{ }          &              &          & &          &              &          & \push{\ket{0}}    & \ghost{M_{ZY}}        & \gate{M_X} \cwx[-1] &
}
\]
\caption{Y measurement implementation using $M_{ZY}$.}
\label{fig:y-measurement-with-multi-pauli-measurement}
\end{figure}

\begin{figure}[t!]
\centering
\[
\Qcircuit @C=.4em @R=.4em {
\lstick{ }          &      &                       & \control[2] \cwx[2]   & \cw                  & \cctrl{1} \cwx[1] \\
\lstick{\ket{\psi}} & /\qw & \multigate{1}{M_{PZ}} & \qw                   & \qw                  & \gate{P}            & \qw & \rstick{e^{-\frac{i\pi P}{8}}\ket{\psi}}     \\
\lstick{T\ket{+}}   & \qw  & \ghost{M_{ZY}}        & \multigate{1}{M_{ZY}} & \gate{M_X}           & \cw \cwx[-1] \\
\lstick{\ket{0}}    & \qw  & \qw                   & \ghost{M_{ZY}}        & \gate{M_{X|Z}}       & \cctrl{-2} \cwx[-2] \\
\lstick{ }          &      & \control[-2] \cwx[-2] & \cw                   & \cctrl{-1} \cwx[-1]
}
\]
\caption{$\frac{\pi}{8}$ rotation implementation. $P$ is a Pauli operator. $M_{PZ}$ and $M_{ZY}$ run in parallel.
         The measurement axis of $M_{X|Z}$ depends on the measurement outcome of $M_{PZ}$.}
\label{fig:pi-over-8-rotation-circuit}
\end{figure}
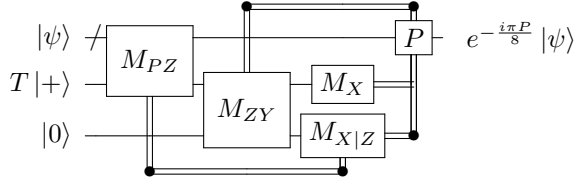

Sequential Pauli-based computation utilizes lattice surgery to implement a multi-qubit Pauli measurement (\autoref{fig:lattice-surgery}).
\autoref{fig:lattice-surgery} (c) uses a twist defect to implement a multi-qubit Pauli measurement involving Y.

$\ket{0}$ and $\ket{+}$ initializations require a single patch and $d$ cycles (\autoref{fig:qubit-initialization} (a)).
Multi-patch and multi-qubit initializations are also possible with $d$ cycles (\autoref{fig:qubit-initialization} (b)).
For $Z$ ($X$) eigenstates, the boundaries corresponding to the logical $Z$ $(X)$ operator can be shortened (\autoref{fig:qubit-initialization} (c)).

Single-qubit $Z$ and $X$ measurements are implemented with transversal $Z$ and $X$ measurements.
A single-qubit $Y$ measurement is implemented with $\ket{0}$ initialization, lattice surgery, and a transversal $Z$ measurement (\autoref{fig:y-measurement-with-multi-pauli-measurement}).
Multi-qubit Pauli measurements are implemented with lattice surgery.
Since transversal $Z$ and $X$ measurements require only one cycle, which is much shorter than $d$ cycles, we neglect this time cost in the following discussions.
With this approximation, single-qubit $Z$ and $X$ measurements require 0 cycles whereas a single-qubit $Y$ measurement and a multi-qubit Pauli measurement require $d$ cycles.

A $\frac{\pi}{8}$ rotation along a Pauli operator $P$ is implemented using multi-Pauli measurements with a magic state, as shown in \autoref{fig:pi-over-8-rotation-circuit}.
This operation requires $d$ cycles.

\subsubsection{Transpilation}
\label{subsec:preliminary-spc-translation}
\begin{figure}[t!]
\centering
\includesvg[width=8.5cm]{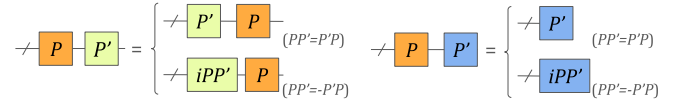}
\caption{Gate translation rules for SPC.
        $P$ and $P'$ are Pauli operators.
        An orange, green, and blue box with $P$ is a $\frac{\pi}{4}$ rotation, $\frac{\pi}{8}$ rotation, and Pauli measurement along $P$, respectively.}
\label{fig:spc-translation-rule}
\end{figure}

A quantum program, given in the form of the quantum IR, is translated into the quantum ISA through the following steps.
First, each arbitrary-angle $Z$ rotation is translated into a sequence of Clifford and $T$ gates using gate synthesis (\autoref{subsec:preliminary-ftqc-compiler}).
Next, each Clifford gate is translated into a sequence of $\frac{\pi}{4}$ rotations.
For example, $H = e^{\frac{i \pi}{4}Z}e^{\frac{i \pi}{4}X}e^{\frac{i \pi}{4}Z}$ and $CX_{12} = e^{\frac{i \pi}{4}Z_1X_2}e^{-\frac{i \pi}{4}Z_1}e^{-\frac{i \pi}{4}X_2}$.
In the final step, $\frac{\pi}{4}$ rotations are moved to the end of the circuit using the rules depicted in \autoref{fig:spc-translation-rule}.
We can ignore the $\frac{\pi}{4}$ rotations at the end of the circuit because they do not affect the measurement outcomes.
The resulting circuit consists of single-qubit initialization, Pauli measurements, and $\frac{\pi}{8}$ rotation gates.
Thus, it is supported by the quantum ISA described in \autoref{subsec:preliminary-spc-isa}.

\begin{figure}[t!]
\centering
\includesvg[width=3cm]{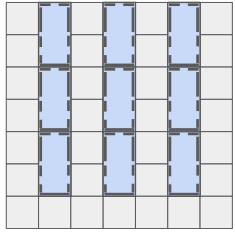}
\caption{
A sequential Pauli-based computation layout with $N = 18$.
Magic state distillation factories are not included.
Each blue rectangle consists of two patches, collectively hosting two data qubits.}
\label{fig:spc-layout}
\end{figure}

Next, we discuss scheduling, mapping, and routing for this compilation scheme.
Through the aforementioned translation, a single-qubit $\frac{\pi}{8}$ $Z$ rotation is translated into a $\frac{\pi}{8}$ rotation along $P$, where $P$ is a multi-qubit Pauli operator.
As a result, performing multiple $\frac{\pi}{8}$ rotations in parallel is challenging, even when they commute.
Hence, sequential Pauli-based computation performs each rotation and measurement sequentially, making scheduling trivial.
This also simplifies mapping and routing, as supporting sequential gate execution is straightforward.
We use the layout depicted in \autoref{fig:spc-layout}, which requires $2N + \sqrt{8N} + 1$ patches, excluding magic state distillation factories, where $N$ is the number of data qubits.

\subsection{Examples of quantum circuits for benchmarking}
In this subsection, we describe two quantum circuits that we use for performance evaluation.

\subsubsection{Random circuit sampling}
\label{subsec:preliminary-rcs}
Random circuit sampling samples from the probability distribution of randomly generated quantum circuits.
Due to the average-case hardness of the task on classical computers~\cite{Bouland2018OnTheComplexity} and the relative ease of experimental realization, it serves as a benchmark for demonstrating quantum supremacy~\cite{Arute2019QuantumSupremacy}.
Random circuit sampling is structured in layers, each of which consists of two steps.
In the first step, a $CZ$ gate is applied to each pair of neighboring qubits.
In the second step, each qubit undergoes a randomly chosen single-qubit gate from $\{S, H, T\}$.

\subsubsection{Hamiltonian simulation with Trotterization}
\label{subsec:preliminary-trotterization}
Hamiltonian simulation is one of the most advantageous tasks in quantum computing.
As a prototypical example, we employ a quantum circuit for simulating the 2D transverse field Ising model using Trotterization.
This is based on~\cite{Beverland2022Assessing, Pearson2020Simulating}.
We consider the time evolution $e^{-iHt}$ of a Hamiltonian
\[
H = -JA + gB = -J\sum_{\ev{j, k}}Z_jZ_k + g\sum_j X_j
\]
for some constants $J$ and $g$.

To approximate $e^{-iHt}$, we split it into $T$ steps, where $T$ is a positive integer.
Observe that $e^{-iHt} = (e^{-iHt/T})^T$.
Each $e^{-iHt/T}$ is approximated using the following fourth-order Trotterization approximation:
\begin{align*}
  U_2(\Delta) &= e^{-iB\Delta/2}e^{-iA\Delta}e^{-iB\Delta/2}, \\
  U_4(\Delta) &= (U_2(\gamma\Delta))^2 U_2((1 - 4\gamma)\Delta)(U_2(\gamma\Delta))^2, \\
  \gamma      &= (4 - 4^{\frac{1}{3}})^{-1}.
\end{align*}
Here, $U_4(\Delta)$ approximates $e^{-iH\Delta}$ up to an error of $\mathcal{O}(\Delta^5)$.
$T$ is chosen to ensure that the total error remains within an acceptable range.
The decomposition yields a sequence of $5T$ $A$ exponentials and $5T + 1$ $B$ exponentials, which can be directly translated into the quantum IR (\autoref{subsec:preliminary-ftqc-compiler}).


\section{Locality-aware Pauli-based computation}
\label{sec:proposal}
In this section, we detail our proposal, locality-aware Pauli-based computation.
In \autoref{subsec:proposal-overview}, we provide an overview of the scheme without going into details.
Next, we list related studies in \autoref{subsec:proposal-related-studies}.
We then examine the validity of our assumption regarding in-place magic state distillation in \autoref{subsec:proposal-inplace-distillation}.
Finally, we present the quantum ISA (\autoref{subsec:proposal-quantum-isa}) and describe the transpilation process (\autoref{subsec:proposal-translation-to-isa}).

\subsection{Overview of the scheme}
\label{subsec:proposal-overview}
We propose locality-aware Pauli-based computation (LAPBC), a new compilation scheme that consists of a transpilation algorithm, a qubit layout, and other compilation algorithms, all of which are designed with an emphasis on locality.
We provide an overview of the transpilation algorithm and qubit layout below.

\begin{figure}[t!]
\centering
\includesvg[width=8.5cm]{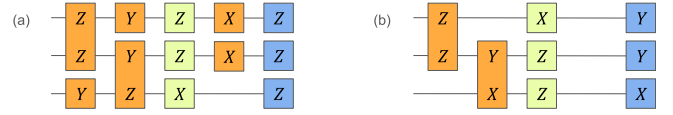}
\caption{Circuit translation of locality-aware Pauli-based computation (LAPBC).
        (a) Original circuit. (b) LAPBC translation result.
        Pauli gates in the circuits are omitted. See \autoref{fig:spc-translation-rule} for the color coding scheme.}
\label{fig:lapbc-translation}
\end{figure}

LAPBC transpiles a quantum IR program (\autoref{subsec:preliminary-ftqc-compiler}) into a sequence of single-qubit $\frac{\pi}{8}$ Pauli rotations, single-qubit or two-qubit $\frac{\pi}{4}$ Pauli rotations, and Pauli measurements (\autoref{fig:lapbc-translation}).
This translation removes single-qubit Clifford gates from the original program and enables efficient program execution similary to SPC.
However, as a crucial difference from SPC, we do not propagate all Clifford gates backward; rather, we restrict ourselves to single-qubit Clifford gates.
\update{This translation preserves the support of each gate, that is, the set of logical qubits that the gate acts on non-trivially.
We refer to this as the preservation of the gate locality of the original program.
As a result, the proposed transpilation allows parallel gate execution.}

\begin{figure}[t!]
\centering
\includesvg[width=8.5cm]{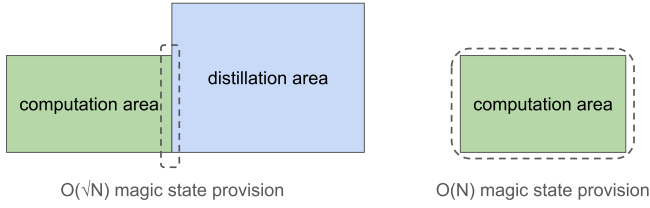}
\caption{Magic state provision in the conventional layout (left) and locality-aware Pauli-based computation (right).}
\label{fig:magic-state-provision}
\end{figure}

\begin{figure}[t!]
\centering
\includesvg[width=7cm]{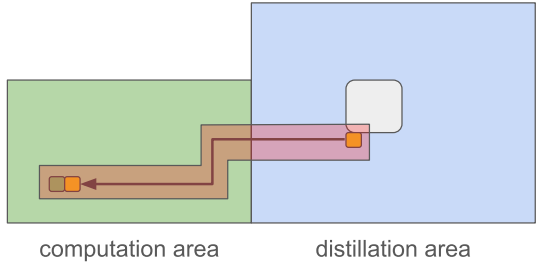}
\caption{A magic state distillation failure (gray) causes a runtime delay that propagates to the magic state transfer path (red).}
\label{fig:runtime-delay-propagation}
\end{figure}

The next most important aspect after LAPBC is that we adopt \textit{factory-less layouts}, which are qubit layouts without separate magic state factories for LAPBC.
Magic state distillation is performed using routing qubits under the assumption that distillation factories are small motivated by recent progress on space efficient magic state distillation protocols~\cite{Itogawa2024Efficient,Gidney2024MagicStateCultivation}.
As discussed in \autoref{sec:introduction}, the conventional layout (\autoref{fig:conventional-layout}) \update{leads to significant magic state transfer costs and} is not well suited for highly parallel algorithms\update{, for the following reasons}.
\update{First,} since a $T$ gate requires a path between the computation and distillation areas, the maximum parallelism is limited by the perimeter of the computation area, which is $\mathcal{O}(\sqrt{N})$, where $N$ is the number of logical data qubits (\autoref{fig:magic-state-provision} (left)).
\update{Second,} the negative impact of distillation failures can be amplified by the conventional layout.
Magic state distillation is a probabilistic process and may fail.
When this happens, the system retries distillation, and any computation dependent on the magic state must wait for its completion.
In the conventional layout, computations involving qubits on the magic state transfer path must also be stalled, even when they are logically independent of the magic state (\autoref{fig:runtime-delay-propagation}), thereby lowering the effective parallelism~\cite{Hirano2024MagicPool}.
Although this issue can be mitigated, doing so introduces additional computational overhead.

A factory-less layout enables local distillation, reduces magic state transfer costs, and allows $\mathcal{O}(N)$ parallel $T$ gate execution (\autoref{fig:magic-state-provision} (right)).
Local distillation also minimizes the runtime delay caused by distillation failures.
By combining the transpilation algorithm and the factory-less layout, locality-aware Pauli-based computation aims to shorten the execution time of highly parallel quantum algorithms.

\subsection{Related studies}
\label{subsec:proposal-related-studies}
SPC~\cite{Litinski2019GameOfSurfaceCodes} is a compilation scheme designed for the conventional layout.
It efficiently performs computations by eliminating Clifford gates and executing multi-qubit Pauli measurements sequentially.
While~\cite{Litinski2019GameOfSurfaceCodes} includes a strategy for parallel computation,~\cite{Chamberland2022Universal} finds this strategy inefficient in practice.

Several resource estimation and compilation studies for the conventional layout support parallel computation~\cite{Beverland2022EdgeDisjointPath, Beverland2022Assessing, Chamberland2022Universal, leblond2024realistic, Molavi2024Dependency, Hamada2024Efficient}.
The limitation on the $T$ parallelism discussed in \autoref{subsec:proposal-overview} applies to these schemes.
For example,~\cite{Beverland2022EdgeDisjointPath} reports $\mathcal{O}(\sqrt{N})$ parallelism, aligning with our analysis.
Some studies, such as~\cite{Beverland2022EdgeDisjointPath, Molavi2024Dependency}, propose mapping, routing, and scheduling algorithms to support parallel computation.
As discussed later in \autoref{subsec:proposal-translation-to-isa}, our compilation scheme can be combined with such algorithms.
Reference~\cite{kobori2024lsqca} proposes a compilation scheme with a layout consisting of a large memory block, a processing unit with a limited size, and distillation factories.
By leveraging memory access locality, this scheme reduces spatial overhead with only a small additional temporal cost.
This is best suited for local and sequential computation, which contrasts with our target applications.

Some studies use alternative layouts.
For example,~\cite{Gidney2021howtofactorbit} uses a layout that does not separate magic state factories from the computation area.
This study is specific to Shor's factoring algorithm, as is the layout.
\update{Reference~\cite{holmes2019resource} proposes placing magic state factories inside the computation area statically with a motivation similar to ours.
An important difference is that our proposal performs magic state distillation using the routing qubits temporarily, rather than relying on statically placed factories.}
Reference~\cite{akahoshi2024compilation} proposes a compilation scheme that performs computation with Clifford gates and arbitrary-angle rotations.
The arbitrary-angle rotations use a state injection protocol performed in the computation area, since it does not rely on magic state distillation.

\subsection{Magic state distillation protocol}
\label{subsec:proposal-inplace-distillation}
In the following discussions, we assume the existence of \textit{in-place} distillation, motivated by recent progress in magic state distillation~\cite{Itogawa2024Efficient, Gidney2024MagicStateCultivation}.
In this subsection, we discuss the validity of this assumption.

Magic state cultivation~\cite{Gidney2024MagicStateCultivation} is a magic state distillation protocol.
The distillation process consists of three stages: injection, cultivation, and escape.
In the injection stage, a magic state is encoded into a color code non-fault-tolerantly.
In the cultivation stage, the magic state is distilled by performing several rounds of Hadamard tests.
At this stage, the magic state is encoded in a color code.
In the escape stage, the magic state is encoded into a grafted code with a higher distance ($d_{g}$) so that we can maintain the fidelity of the magic state with error correction.
The support of the grafted code is mostly square-shaped, which is attractive for in-place distillation.

Reference~\cite{Gidney2024MagicStateCultivation} reports a logical error rate of $2 \times 10^{-9}$ for the resultant magic state, with a physical error rate of $10^{-3}$ and $d_{g} = 15$.
This error rate is comparable to that of the surface code with $d = 15$, which is approximately $10^{-9}$.
Moreover, in many cases, the error rate of magic states can be higher than that of data and routing qubits.
For example, random circuit sampling (\autoref{subsec:preliminary-rcs}) performs $CZ$, $H$, $S$, and $T$ gates.
This suggests that magic states are used in approximately 1/6 of the gates.
If there is one routing qubit for each data qubit, the error rate of a magic state can be 12 times higher than that of a data qubit or a routing qubit.
This difference can be leveraged to reduce $d_g$ and/or increase the acceptance ratio.
This rough estimation suggests that magic state cultivation offers in-place distillation within a certain fidelity range.
Although this range is limited, future improvements in distillation protocols will expand the applicability of in-place distillation.

In this study, we assume that distillation is performed in a single patch to simplify the discussion.
However, locality-aware Pauli-based computation is compatible with a distillation protocol that uses a few patches, though this would require increased complexity in mapping, routing, and scheduling algorithms.
Whether such a setting improves applicability by reducing the error rate and increasing the acceptance ratio remains an open question.

\subsection{Quantum ISA and its implementation}
\label{subsec:proposal-quantum-isa}
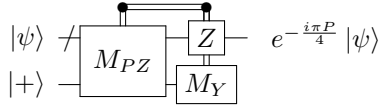
\begin{figure}[t!]
\centering
\[
\Qcircuit @C=.4em @R=.4em {
\lstick{ }          &      & \control[1] \cwx[1]   &  \cctrl{1} \cwx[1] \\
\lstick{\ket{\psi}} & /\qw & \multigate{1}{M_{PZ}} &  \gate{Z}            & \qw & \rstick{e^{-\frac{i\pi P}{4}}\ket{\psi}}     \\
\lstick{\ket{+}}    & \qw  & \ghost{M_{PZ}}        &  \gate{M_Y} \cwx[-1]
}
\]
\caption{$\frac{\pi}{4}$ rotation implementation. $P$ is a Pauli operator.}
\label{fig:lapbc-pi-over-4-rotation-circuit}
\end{figure}

\begin{figure}[t!]
\centering
\includesvg[width=8cm]{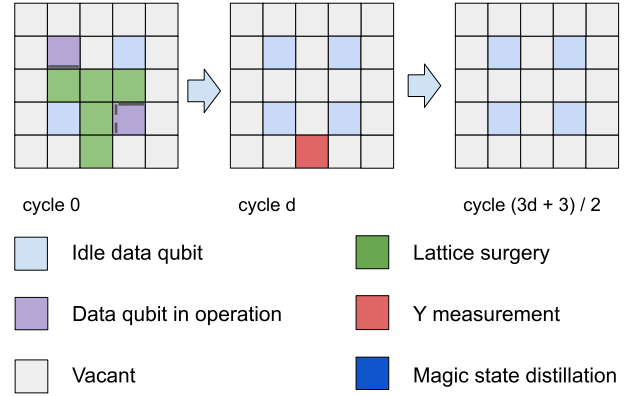}
\caption{$\frac{\pi}{4}$ rotation along $ZY$.}
\label{fig:lapbc-pi-over-4-yz-rotation}
\end{figure}

\begin{figure}[t!]
\centering
\[
\Qcircuit @C=.4em @R=.4em {
\lstick{\ket{\psi}} & \qw & \multigate{1}{M_{PZ}} & \qw                 & \gate{P}            & \qw & \rstick{e^{-\frac{i\pi P}{8}}\ket{\psi}}     \\
\lstick{T\ket{+}}   & \qw & \ghost{M_{PZ}}        & \gate{M_{X|Y}}      & \cctrl{-1} \cwx[-1] & \\
\lstick{ }          &     & \control[-1] \cwx[-1] & \cctrl{-1} \cwx[-1] 
}
\]
\caption{Single-qubit $\frac{\pi}{8}$ rotation implementation. $P$ is a Pauli operator.
         The measurement axis of $M_{X|Y}$ depends on the measurement outcome of $M_{PZ}$.}
\label{fig:lapbc-pi-over-8-rotation-circuit}
\end{figure}
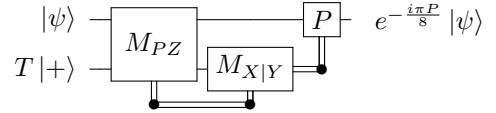

\begin{figure}[t!]
\centering
\includesvg[width=5.5cm]{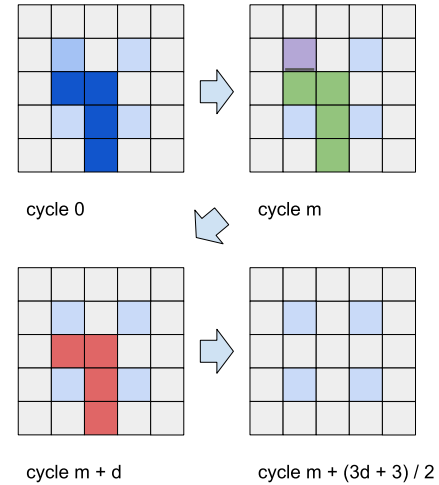}
\caption{$\frac{\pi}{8}$ rotation along $Z$.
         See \autoref{fig:lapbc-pi-over-4-yz-rotation} for the coloring convention.}
\label{fig:lapbc-pi-over-8-z-rotation}
\end{figure}

The quantum ISA for locality-aware Pauli-based computation consists of the following operations:
\begin{enumerate}
 \item Single-qubit $\ket{0}$ initialization,
 \item Single-qubit Pauli measurement,
 \item Multi-qubit Pauli measurement,
 \item $\frac{\pi}{4}$ rotation along a two-qubit Pauli operator, and
 \item $\frac{\pi}{8}$ rotation along a single-qubit Pauli operator.
\end{enumerate}
The first three operations are the same as those in SPC (\autoref{subsec:preliminary-spc-isa}), and we use the same implementations except for the single-qubit $Y$ measurement, which we replace with an in-place $Y$ measurement~\cite{gidney2024inplace} that takes $\frac{d + 3}{2}$ cycles.
A $\frac{\pi}{4}$ rotation along a two-qubit Pauli operator $P$ is implemented using multi-qubit Pauli measurements, as shown in \autoref{fig:lapbc-pi-over-4-rotation-circuit}.
This operation takes $\frac{3d + 3}{2}$ cycles.
\autoref{fig:lapbc-pi-over-4-yz-rotation} depicts the computation process for a $\frac{\pi}{4}$ rotation along $ZY$.
A $\frac{\pi}{8}$ rotation along a single-qubit Pauli operator $P$ is implemented using multi-qubit Pauli measurements, as shown in \autoref{fig:lapbc-pi-over-8-rotation-circuit}.
This operation requires $m + \frac{3d + 3}{2}$ cycles, where $m$ is the time required for magic state distillation.
\autoref{fig:lapbc-pi-over-8-z-rotation} depicts the computation process for a $\frac{\pi}{8}$ rotation along $Z$.
In the figure, four magic state distillation processes are performed to mitigate the low success probability of distillation.
If at least one of the four distillation processes succeeds, the $\frac{\pi}{8}$ rotation completes on time.
Otherwise, the system retries distillation, leading to a \textit{runtime delay}.

\update{When performing a multi-qubit Pauli measurement used in $\frac{\pi}{4}$ and $\frac{\pi}{8}$ rotations, we need to choose the correct edges of the target data qubit patches based on the rotation axis, as discussed in \autoref{subsec:preliminary-spc}.
In all examples in this subsection, a horizontal (vertical) edge of a logical data qubit patch corresponds to its logical $Z$ ($X$) operator.
Hence, for instance, to perform a $\frac{\pi}{4}$ rotation along $XX$, each target data qubit patch must share one of its vertical edges with the lattice surgery ancilla patches.
This allows us to ensure that each $\frac{\pi}{4}$ rotation requires a fixed number of cycles, independent of the rotation axis.
The same applies to $\frac{\pi}{8}$ rotations, except in cases of magic state distillation failure.
Note that magic states can be prepared with arbitrary orientation, which relaxes routing constraints for $\frac{\pi}{8}$ rotations.}

\subsection{Transpilation to the quantum ISA}
\label{subsec:proposal-translation-to-isa}
\begin{figure}[t!]
\centering
\includesvg[width=8.5cm]{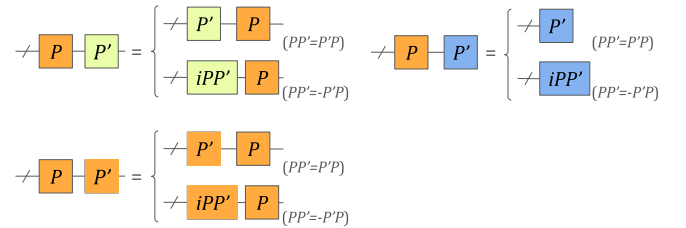}
\caption{Gate translation rules for locality-aware Pauli-based computation.
        $P$ is a single-qubit Pauli operator and $P'$ is a Pauli operator.
        See \autoref{fig:spc-translation-rule} for the coloring convention.}
\label{fig:lapbc-translation-rule}
\end{figure}

A quantum program given in the form of the quantum IR is transpiled into the quantum ISA using the following steps.
First, each arbitrary angle $Z$ rotation is translated into a sequence of Clifford and $T$ gates using gate synthesis (\autoref{subsec:preliminary-ftqc-compiler}).
Next, each Clifford gate is translated into a sequence of $\frac{\pi}{4}$ rotations.
Finally, single-qubit $\frac{\pi}{4}$ rotations are moved to the end of the circuit, using the rules illustrated in \autoref{fig:lapbc-translation-rule}.

\begin{figure}[t!]
\centering
\includesvg[width=5.5cm]{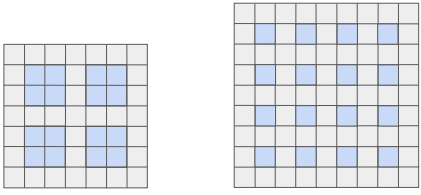}
\caption{The standard layout (left) and sparse layout (right), each containing 16 logical data qubits.}
\label{fig:lapbc-layouts}
\end{figure}

We now discuss mapping, routing, and scheduling for this compilation scheme.
Locality-aware Pauli-based computation is compatible with any qubit layout that supports the quantum ISA implementation, including the examples shown in \autoref{fig:lapbc-pi-over-4-yz-rotation} and \autoref{fig:lapbc-pi-over-8-z-rotation}.
In this study, we consider two layouts, \textit{standard} and \textit{sparse} (\autoref{fig:lapbc-layouts}).
The standard and sparse layouts require approximately $2.25N$ and $4N$ patches, respectively, where $N$ is the number of data qubits.
Unlike SPC, efficient mapping, routing, and scheduling are crucial for locality-aware Pauli-based computation, as it aims to perform many gates simultaneously by leveraging locality.
In \autoref{sec:performance-evaluation}, we implement basic mapping, routing, and scheduling strategies in our scheduler.
We expect that locality-aware Pauli-based computation will perform better with more sophisticated algorithms such as those proposed in~\cite{Beverland2022EdgeDisjointPath, Molavi2024Dependency}, but we leave a detailed investigation for future work.

\section{Performance evaluation}
\label{sec:performance-evaluation}
This section presents a performance evaluation of locality-aware Pauli-based computation.
We begin by describing the design and implementation of the scheduler and runtime simulator in \autoref{subsec:performance-evaluation-scheduler} and \autoref{subsec:performance-evaluation-simulator}.
We then describe the simulation parameters in \autoref{subsec:performance-evaluation-parameter-settings}.
Finally, we present the simulation results in \autoref{subsec:performance-evaluation-results}.
The scheduler and runtime simulator is publicly available~\cite{Hirano2025LAPBCSimulation}.

\subsection{Scheduler implementation}
\label{subsec:performance-evaluation-scheduler}
\begin{figure}[t!]
\centering
\includesvg[width=7cm]{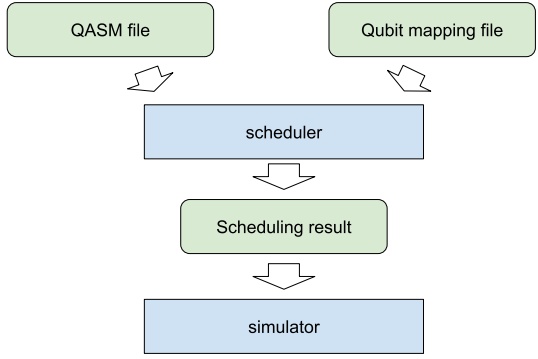}
\caption{System configuration of the scheduler and simulator used in the performance evaluation.}
\label{fig:simulation-system-configuration-diagram}
\end{figure}

\autoref{fig:simulation-system-configuration-diagram} shows the system configuration used in the simulation.
The scheduler outputs the scheduling result, given a QASM program and qubit mapping file.
It implements gate synthesis, mapping, routing, and scheduling algorithms, which we describe below.

\textbf{Gate synthesis: } For the simulation, we emulate the mixed diagonal algorithm by replacing an arbitrary-angle $Z$ rotation $e^{i{\theta}Z}$ with a sequence of $l$ randomly chosen $\frac{\pi}{8}$ rotations, followed by two randomly chosen $\frac{\pi}{4}$ rotations.
The value of $l$ is sampled from a normal distribution with mean $1.5\delta$, where $2^{-\delta}$ is the target precision of the rotation angle.

\textbf{Mapping: } As shown in \autoref{fig:simulation-system-configuration-diagram}, the mapping configuration is provided externally to the scheduler to simplify its implementation.

\textbf{Routing: } To execute a two-qubit $\frac{\pi}{4}$ rotation, we find a path between the two patches using a breadth-first search algorithm.
Ancilla patches near the data qubits may be required depending on the rotation axis.
An example is shown in \autoref{fig:lapbc-pi-over-4-yz-rotation}.
Additionally, one ancilla patch is required for a $Y$ measurement.

For a single-qubit $\frac{\pi}{8}$ rotation, we allocate $D$ connected patches for magic state distillation, where $D$ is a simulator parameter.
The allocation is computed using a breadth-first search algorithm to minimize the diameter of the distillation area.
Similar to the two-qubit $\frac{\pi}{4}$ case, ancilla patches near the data qubits may be required depending on the rotation axis.

\begin{figure}[t!]
\centering
\includegraphics[width=6cm]{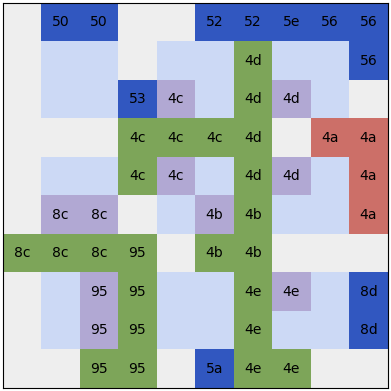}
\caption{Schedule for $6\times6$ random circuit sampling.
         Each number represents an instruction ID.
         See \autoref{fig:lapbc-pi-over-4-yz-rotation} for the color coding scheme.}
\label{fig:scheduling-result}
\end{figure}

\textbf{Scheduling: } We use a greedy algorithm (\autoref{alg:schedule}) to schedule operations.
The algorithm takes as input a sequence of quantum ISA instructions, denoted $instructions$, ordered by gate dependencies.
It produces a schedule consisting of \textit{microinstructions} assigned to each patch and cycle.
Each microinstruction falls into one of the following categories: \textit{idle data qubit}, \textit{data qubit in operation}, \textit{vacant}, \textit{lattice surgery}, \textit{Y measurement}, or \textit{magic state distillation}, as indicated by the color coding in \autoref{fig:lapbc-pi-over-4-yz-rotation}.
\autoref{fig:scheduling-result} shows a snapshot of the schedule for $6 \times 6$ random circuit sampling at a specific cycle.

\begin{figure}[!t]
\begin{algorithm}[H]
\caption{Scheduling algorithm.}
\label{alg:schedule}
\begin{algorithmic}[1]
 \FOR{$i$ in $instructions$}
  \STATE{$support$ = support qubits of $i$}
  \STATE{$cycle$ = maximal scheduled cycles on $support$}
  \WHILE{true}
   \IF {routing for $op$ is possible at $cycle$}
    \STATE{Commit the schedule for $i$ starting at $cycle$.}
    \STATE{\textbf{break}}
   \ENDIF
   \STATE $cycle$ += 1
  \ENDWHILE
 \ENDFOR
\end{algorithmic}
\end{algorithm}
\end{figure}

\subsection{Simulator implementation}
\label{subsec:performance-evaluation-simulator}
The runtime simulator estimates the execution time, based on a schedule, as depicted in \autoref{fig:simulation-system-configuration-diagram}.
Most of the simulation is straightforward; the only non-trivial component is the calculation of \textit{runtime delay}.
Magic state distillation is the only source of runtime delay.
In this simulation, we model the time cost of magic state distillation as $m \cdot G(p)$, where $m$ is the time required for a single distillation round, and $G(p)$
is a geometric random variable with success probability $p$, sampled independently for each event.

As discussed in \autoref{subsec:proposal-overview}, runtime delay can propagate.
The simulation uses the following runtime delay propagation rules.
\begin{itemize}
 \item Any microinstruction on a patch except for \textit{idle data qubit} and \textit{vacant} must wait for all preceding microinstructions on the same patch to complete.
 \item \textit{Lattice surgery} and \textit{data qubit in operation} microinstructions must wait for all preceding microinstructions on the involved patches to complete.
\end{itemize}

\subsection{Parameter settings}
\label{subsec:performance-evaluation-parameter-settings}
\autoref{tab:simulation-parameter-settings} shows the parameters we use in the simulations that follow.
We use the same code distance, $d = 15$, across all simulations.
In general, larger problem sizes require larger code distances than smaller ones.
In this study, however, we use the same code distance for the following reasons.
First, the differences in problem size are relatively small in our case.
With the physical error rate $10^{-3}$ and surface code threshold $10^{-2}$, increasing code distance by 2 corresponds to increasing the problem size by 10 times in qubitcycles.
In comparison, the 12$\times$12 instance is only four times larger than the 6$\times$6 instance.
Second, comparing results with the same code distance is significantly easier than comparing those with different code distances.
We apply the same reasoning to the use of common values for $m$, $p_{\textrm{success}}$, and $\rho$ across all simulations.
Note also that the code distance $d = 15$ provides sufficient fidelity for all simulations performed below.

\begin{table}[t!]
  \caption{Parameter settings used in the simulations.}
  \centering
  \begin{tabular}{r|c|c}
    symbol                 & description                         & value \\
    \hline\hline
    $d$                    & code distance                       & 15 \\
    $m$                    & distillation time cost              & 27 \\
    $p_{\textrm{success}}$ & distillation acceptance ratio       & 0.25 \\
    $\rho$                 & arbitrary-angle rotation precision  & $10^{-7}$ \\
    \hline
  \end{tabular}
  \label{tab:simulation-parameter-settings}
\end{table}

\subsection{Results}
\label{subsec:performance-evaluation-results}
\begin{figure}[t!]
\centering
\includesvg[width=8.5cm]{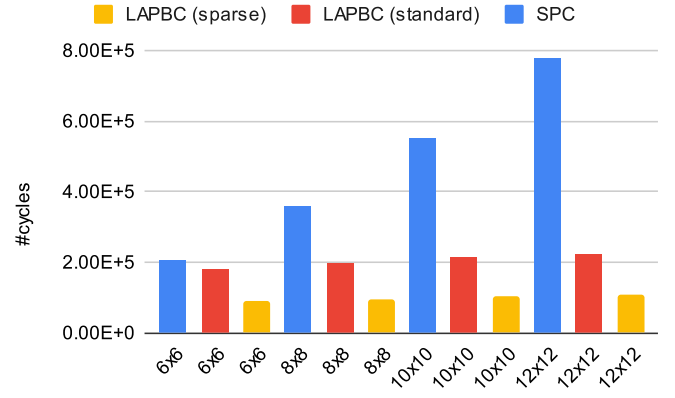}
\caption{Simulation results for random circuit sampling with 500 layers.}
\label{fig:rcs-simulation-result}
\end{figure}

\begin{figure}[t!]
\centering
\includesvg[width=8.5cm]{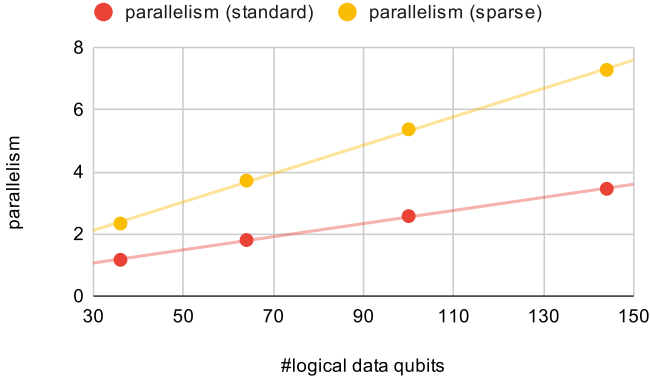}
\caption{Parallelism of locality-aware Pauli-based computation with random circuit sampling.}
\label{fig:rcs-parallelism}
\end{figure}

\begin{figure}[t!]
\centering
\includesvg[width=8.5cm]{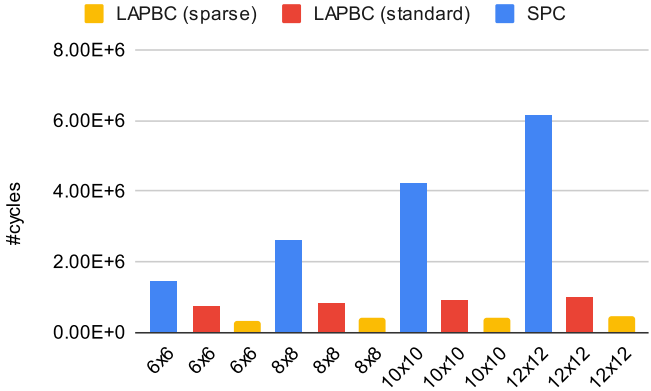}
\caption{Simulation results for Hamiltonian simulation with Trotterization.}
\label{fig:trotter-simulation-result}
\end{figure}

We performed simulations of random circuit sampling (\autoref{subsec:preliminary-rcs}) and the Hamiltonian simulation with Trotterization (\autoref{subsec:preliminary-trotterization}).
\update{These applications have natural 2D structures, and we performed manual qubit mapping based on these structures.}
\autoref{fig:rcs-simulation-result} shows the simulation results for random circuit sampling with 500 layers.
While the cycle count for SPC grows approximately proportionally with the number of logical data qubits, that for locality-aware Pauli-based computation—using both standard and sparse layouts—increases more gradually due to greater gate-level parallelism.
Note that the results for locality-aware Pauli-based computation account for runtime delays from magic state distillation, whereas the results for SPC do not.
This is because runtime delays are easier to mitigate for SPC due to its sequential nature and the small spatial cost of magic state distillation.
\autoref{fig:rcs-parallelism} shows the parallelism of locality-aware Pauli-based computation, where parallelism is defined as the number of cycles required for SPC divided by the number of cycles required for locality-aware Pauli-based computation.
Parallelism scales linearly with the number of data qubits $N$, following a trend of the form $aN + b$ for some constants $a$ and $b$, and surpasses the $\mathcal{O}(\sqrt{N})$ scaling of the conventional layout (\autoref{subsec:proposal-overview}).
\autoref{fig:trotter-simulation-result} shows the Hamiltonian simulation results, which exhibit a similar trend.

For 6$\times$6 (12$\times$12) random circuit sampling, locality-aware Pauli-based computation with the standard layout achieves a 14\% (71\%) reduction in execution time relative to SPC.
Similarly, for 6$\times$6 (12$\times$12) Hamiltonian simulation with Trotterization, it achieves a 48\% (84\%) reduction.
The spatial overhead of the standard layout is comparable to that of SPC, so these improvements are obtained with little or no additional spatial cost.
For 6$\times$6 (12$\times$12) random circuit sampling, the sparse layout achieves a 57\% (86\%) reduction in execution time relative to SPC.
For Hamiltonian simulation, the reductions are 76\% and 93\%, respectively.
The sparse layout incurs approximately twice the spatial overhead of SPC.

\begin{figure}[t!]
\centering
\includesvg[width=8.5cm]{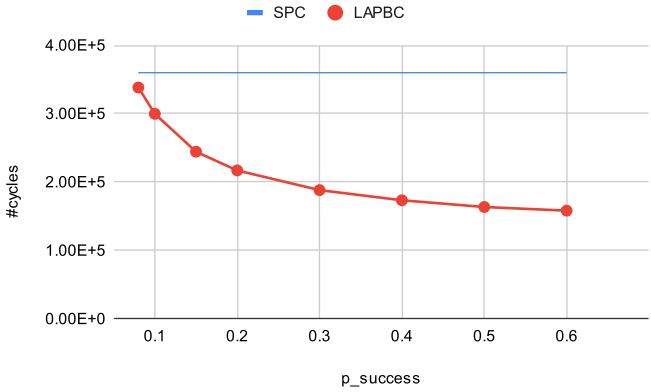}
\caption{Simulation results for 8$\times$8 random circuit sampling with varying $p_{\textrm{success}}$.}
\label{fig:rcs-simulation-result-with-various-p-success}
\end{figure}

\autoref{fig:rcs-simulation-result-with-various-p-success} shows the simulation results for $8 \times 8$ random circuit sampling using the standard layout, with varying values of $p_{\textrm{success}}$.
The results demonstrate that the cycle count is sensitive to $p_{\textrm{success}}$, indicating that the distillation cost remains a significant factor for highly parallel algorithms.
Moreover, in this case, increasing $p_{\textrm{success}}$ up to 0.4 significantly reduces the overall time cost, whereas further increases have a negligible impact.

\section{Conclusion}
\label{sec:conclusion}
In this study, we proposed locality-aware Pauli-based computation, a novel compilation scheme optimized for highly parallel quantum algorithms. 
By preserving gate locality in the original algorithm and performing magic state distillation locally, the scheme enables the parallel execution of many $T$ gates.

Numerical simulations in \autoref{sec:performance-evaluation} show that, for 6$\times$6 (12$\times$12) random circuit sampling, locality-aware Pauli-based computation achieves a 14\% (71\%) reduction in execution time relative to SPC, with little or no additional spatial overhead.
Similarly, for 6$\times$6 (12$\times$12) Hamiltonian simulation with Trotterization, it achieves a 48\% (84\%) reduction.
Even greater reductions can be achieved with increased spatial overhead, as demonstrated by the sparse layout.
These results demonstrate that locality-aware Pauli-based computation scales favorably in execution time as the number of qubits increases.
Therefore, we conclude that locality-aware Pauli-based computation will become a standard compilation scheme targeting highly parallel quantum algorithms when high-performance magic state distillation is available.

Our scheduler and runtime simulator reveal the relationship between the magic state distillation cost---including the acceptance ratio---and the overall computational cost for a given quantum algorithm.  
This provides useful design guidelines for developing magic state distillation protocols.

When magic state distillation was hundreds of times more expensive than Clifford gates, it was reasonable to treat the problems of magic state preparation and consumption separately. 
However, as the cost of distillation continues to decrease, considering preparation and consumption jointly becomes increasingly important.
This study highlights the effectiveness of such an integrated approach in the design of compilation schemes.

\update{The mapping and routing algorithms implemented for the numerical simulations are simple.
We expect that locality-aware Pauli-based computation will achieve better performance with more sophisticated algorithms such as those proposed in~\cite{Beverland2022EdgeDisjointPath, Molavi2024Dependency}, but we leave a detailed investigation to future work.
Integrating the factory-less layout with the direct Clifford+$T$ compilation~\cite{leblond2025quantum} presents another interesting direction.}

In this work, we assume that each distillation is performed within a single patch. 
However, locality-aware Pauli-based computation is also compatible with distillation protocols that use multiple patches, though this would introduce additional complexity.
Since routing qubits are temporarily used for distillation, the number of patches required by a distillation factory is a more important factor than the total number of physical qubits it occupies.
Whether such a setting is more effective in practice—by reducing the logical error rate or increasing the acceptance ratio—remains an important open question.

\section*{Acknowledgment}
This work is supported by MEXT Quantum Leap Flagship Program (MEXT Q-LEAP) Grant No. JP- MXS0118067394 and JPMXS0120319794, JST COI- NEXT Grant No. JPMJPF2014, and JST Moonshot R\&D Grant No. JPMJMS2061.
YH thanks Riki Toshio and Yutaro Akahoshi for early discussions.

\clearpage
\bibliographystyle{IEEETran}
\bibliography{refs}

\begin{thebibliography}{10}
\providecommand{\url}[1]{#1}
\csname url@samestyle\endcsname
\providecommand{\newblock}{\relax}
\providecommand{\bibinfo}[2]{#2}
\providecommand{\BIBentrySTDinterwordspacing}{\spaceskip=0pt\relax}
\providecommand{\BIBentryALTinterwordstretchfactor}{4}
\providecommand{\BIBentryALTinterwordspacing}{\spaceskip=\fontdimen2\font plus
\BIBentryALTinterwordstretchfactor\fontdimen3\font minus \fontdimen4\font\relax}
\providecommand{\BIBforeignlanguage}[2]{{%
\expandafter\ifx\csname l@#1\endcsname\relax
\typeout{** WARNING: IEEEtran.bst: No hyphenation pattern has been}%
\typeout{** loaded for the language `#1'. Using the pattern for}%
\typeout{** the default language instead.}%
\else
\language=\csname l@#1\endcsname
\fi
#2}}
\providecommand{\BIBdecl}{\relax}
\BIBdecl

\bibitem{Abrams1999Quantum}
\BIBentryALTinterwordspacing
D.~S. Abrams and S.~Lloyd, ``Quantum algorithm providing exponential speed increase for finding eigenvalues and eigenvectors,'' \emph{Phys. Rev. Lett.}, vol.~83, pp. 5162--5165, Dec 1999. [Online]. Available: \url{https://link.aps.org/doi/10.1103/PhysRevLett.83.5162}
\BIBentrySTDinterwordspacing

\bibitem{Shor1994Factoring}
P.~W. Shor, ``Algorithms for quantum computation: discrete logarithms and factoring,'' \emph{Proceedings 35th Annual Symposium on Foundations of Computer Science}, pp. 124--134, 1994.

\bibitem{Kitaev2003}
\BIBentryALTinterwordspacing
A.~Kitaev, ``Fault-tolerant quantum computation by anyons,'' \emph{Annals of Physics}, vol. 303, no.~1, p. 2–30, Jan. 2003. [Online]. Available: \url{http://dx.doi.org/10.1016/S0003-4916(02)00018-0}
\BIBentrySTDinterwordspacing

\bibitem{Bravyi1998}
S.~B. Bravyi and A.~Y. Kitaev, ``Quantum codes on a lattice with boundary,'' 1998.

\bibitem{Arya2024QuantumErrorCorrection}
\BIBentryALTinterwordspacing
G.~Q. AI and Collaborators, ``Quantum error correction below the surface code threshold,'' \emph{Nature}, 2024. [Online]. Available: \url{https://doi.org/10.1038/s41586-024-08449-y}
\BIBentrySTDinterwordspacing

\bibitem{Bravyi2005}
\BIBentryALTinterwordspacing
S.~Bravyi and A.~Kitaev, ``Universal quantum computation with ideal clifford gates and noisy ancillas,'' \emph{Phys. Rev. A}, vol.~71, p. 022316, Feb 2005. [Online]. Available: \url{https://link.aps.org/doi/10.1103/PhysRevA.71.022316}
\BIBentrySTDinterwordspacing

\bibitem{Litinski2019GameOfSurfaceCodes}
\BIBentryALTinterwordspacing
D.~Litinski, ``A {G}ame of {S}urface {C}odes: {L}arge-{S}cale {Q}uantum {C}omputing with {L}attice {S}urgery,'' \emph{{Quantum}}, vol.~3, p. 128, Mar. 2019. [Online]. Available: \url{https://doi.org/10.22331/q-2019-03-05-128}
\BIBentrySTDinterwordspacing

\bibitem{Beverland2022EdgeDisjointPath}
\BIBentryALTinterwordspacing
M.~Beverland, V.~Kliuchnikov, and E.~Schoute, ``Surface code compilation via edge-disjoint paths,'' \emph{PRX Quantum}, vol.~3, p. 020342, May 2022. [Online]. Available: \url{https://link.aps.org/doi/10.1103/PRXQuantum.3.020342}
\BIBentrySTDinterwordspacing

\bibitem{Beverland2022Assessing}
\BIBentryALTinterwordspacing
M.~E. Beverland, P.~Murali, M.~Troyer, K.~M. Svore, T.~Hoefler, V.~Kliuchnikov, G.~H. Low, M.~Soeken, A.~Sundaram, and A.~Vaschillo, ``Assessing requirements to scale to practical quantum advantage,'' 2022. [Online]. Available: \url{https://arxiv.org/abs/2211.07629}
\BIBentrySTDinterwordspacing

\bibitem{Chamberland2022Universal}
\BIBentryALTinterwordspacing
C.~Chamberland and E.~T. Campbell, ``Universal quantum computing with twist-free and temporally encoded lattice surgery,'' \emph{PRX Quantum}, vol.~3, p. 010331, Feb 2022. [Online]. Available: \url{https://link.aps.org/doi/10.1103/PRXQuantum.3.010331}
\BIBentrySTDinterwordspacing

\bibitem{leblond2024realistic}
T.~LeBlond, C.~Dean, G.~Watkins, and R.~Bennink, ``Realistic cost to execute practical quantum circuits using direct clifford+ t lattice surgery compilation,'' \emph{ACM Transactions on Quantum Computing}, vol.~5, no.~4, pp. 1--28, 2024.

\bibitem{Molavi2024Dependency}
\BIBentryALTinterwordspacing
A.~Molavi, A.~Xu, S.~Tannu, and A.~Albarghouthi, ``Dependency-aware compilation for surface code quantum architectures,'' 2024. [Online]. Available: \url{https://arxiv.org/abs/2311.18042}
\BIBentrySTDinterwordspacing

\bibitem{Hamada2024Efficient}
\BIBentryALTinterwordspacing
K.~Hamada, Y.~Suzuki, and Y.~Tokunaga, ``Efficient and high-performance routing of lattice-surgery paths on three-dimensional lattice,'' 2024. [Online]. Available: \url{https://arxiv.org/abs/2401.15829}
\BIBentrySTDinterwordspacing

\bibitem{Goto2016Minimizing}
H.~Goto, ``Minimizing resource overheads for fault-tolerant preparation of encoded states of the steane code,'' \emph{Scientific Reports volume 6, Article number: 19578}, 2016.

\bibitem{Litinski2019magicstate}
\BIBentryALTinterwordspacing
D.~Litinski, ``Magic {S}tate {D}istillation: {N}ot as {C}ostly as {Y}ou {T}hink,'' \emph{{Quantum}}, vol.~3, p. 205, Dec. 2019. [Online]. Available: \url{https://doi.org/10.22331/q-2019-12-02-205}
\BIBentrySTDinterwordspacing

\bibitem{Itogawa2024Efficient}
T.~Itogawa, Y.~Takada, Y.~Hirano, and K.~Fujii, ``Even more efficient magic state distillation by zero-level distillation,'' 2024.

\bibitem{Hirano2024Leveraging}
Y.~Hirano, T.~Itogawa, and K.~Fujii, ``Leveraging zero-level distillation to generate high-fidelity magic states,'' in \emph{2024 IEEE International Conference on Quantum Computing and Engineering (QCE)}, vol.~1.\hskip 1em plus 0.5em minus 0.4em\relax IEEE, 2024, pp. 843--853.

\bibitem{Smith2024Mitigating}
S.~Smith, B.~Brown, and S.~Bartlett, ``Mitigating errors in logical qubits,'' \emph{Communications Physics}, vol.~7, 11 2024.

\bibitem{Gidney2024MagicStateCultivation}
\BIBentryALTinterwordspacing
C.~Gidney, N.~Shutty, and C.~Jones, ``Magic state cultivation: growing t states as cheap as cnot gates,'' 2024. [Online]. Available: \url{https://arxiv.org/abs/2409.17595}
\BIBentrySTDinterwordspacing

\bibitem{riesebos2017pauli}
L.~Riesebos, X.~Fu, S.~Varsamopoulos, C.~G. Almudever, and K.~Bertels, ``Pauli frames for quantum computer architectures,'' in \emph{Proceedings of the 54th Annual Design Automation Conference 2017}, 2017, pp. 1--6.

\bibitem{Horsman2012}
\BIBentryALTinterwordspacing
D.~Horsman, A.~G. Fowler, S.~Devitt, and R.~V. Meter, ``Surface code quantum computing by lattice surgery,'' \emph{New Journal of Physics}, vol.~14, no.~12, p. 123011, dec 2012. [Online]. Available: \url{https://dx.doi.org/10.1088/1367-2630/14/12/123011}
\BIBentrySTDinterwordspacing

\bibitem{Svore2018QSharp}
\BIBentryALTinterwordspacing
K.~Svore, A.~Geller, M.~Troyer, J.~Azariah, C.~Granade, B.~Heim, V.~Kliuchnikov, M.~Mykhailova, A.~Paz, and M.~Roetteler, ``Q\#: Enabling scalable quantum computing and development with a high-level dsl,'' in \emph{Proceedings of the Real World Domain Specific Languages Workshop 2018}, ser. RWDSL2018.\hskip 1em plus 0.5em minus 0.4em\relax New York, NY, USA: Association for Computing Machinery, 2018. [Online]. Available: \url{https://doi.org/10.1145/3183895.3183901}
\BIBentrySTDinterwordspacing

\bibitem{Green2013Quipper}
\BIBentryALTinterwordspacing
A.~S. Green, P.~L. Lumsdaine, N.~J. Ross, P.~Selinger, and B.~Valiron, ``Quipper: a scalable quantum programming language,'' \emph{SIGPLAN Not.}, vol.~48, no.~6, p. 333–342, Jun. 2013. [Online]. Available: \url{https://doi.org/10.1145/2499370.2462177}
\BIBentrySTDinterwordspacing

\bibitem{QIRSpec2021}
\BIBentryALTinterwordspacing
{QIR Alliance}, \emph{{QIR Specification}}, 2021, also see \url{https://qir-alliance.org}. [Online]. Available: \url{https://github.com/qir-alliance/qir-spec}
\BIBentrySTDinterwordspacing

\bibitem{Kliuchnikov2023Shorter}
\BIBentryALTinterwordspacing
V.~Kliuchnikov, K.~Lauter, R.~Minko, A.~Paetznick, and C.~Petit, ``Shorter quantum circuits via single-qubit gate approximation,'' \emph{Quantum}, vol.~7, p. 1208, Dec. 2023. [Online]. Available: \url{http://dx.doi.org/10.22331/q-2023-12-18-1208}
\BIBentrySTDinterwordspacing

\bibitem{Bouland2018OnTheComplexity}
\BIBentryALTinterwordspacing
A.~Bouland, B.~Fefferman, C.~Nirkhe, and U.~V. Vazirani, ``On the complexity and verification of quantum random circuit sampling,'' \emph{Nature Physics}, vol.~15, pp. 159--163, 2018. [Online]. Available: \url{https://api.semanticscholar.org/CorpusID:125264133}
\BIBentrySTDinterwordspacing

\bibitem{Arute2019QuantumSupremacy}
\BIBentryALTinterwordspacing
F.~Arute, K.~Arya, R.~Babbush, D.~Bacon, J.~Bardin, R.~Barends, R.~Biswas, S.~Boixo, F.~Brandao, D.~Buell, B.~Burkett, Y.~Chen, J.~Chen, B.~Chiaro, R.~Collins, W.~Courtney, A.~Dunsworth, E.~Farhi, B.~Foxen, A.~Fowler, C.~M. Gidney, M.~Giustina, R.~Graff, K.~Guerin, S.~Habegger, M.~Harrigan, M.~Hartmann, A.~Ho, M.~R. Hoffmann, T.~Huang, T.~Humble, S.~Isakov, E.~Jeffrey, Z.~Jiang, D.~Kafri, K.~Kechedzhi, J.~Kelly, P.~Klimov, S.~Knysh, A.~Korotkov, F.~Kostritsa, D.~Landhuis, M.~Lindmark, E.~Lucero, D.~Lyakh, S.~Mandrà, J.~R. McClean, M.~McEwen, A.~Megrant, X.~Mi, K.~Michielsen, M.~Mohseni, J.~Mutus, O.~Naaman, M.~Neeley, C.~Neill, M.~Y. Niu, E.~Ostby, A.~Petukhov, J.~Platt, C.~Quintana, E.~G. Rieffel, P.~Roushan, N.~Rubin, D.~Sank, K.~J. Satzinger, V.~Smelyanskiy, K.~J. Sung, M.~Trevithick, A.~Vainsencher, B.~Villalonga, T.~White, Z.~J. Yao, P.~Yeh, A.~Zalcman, H.~Neven, and J.~Martinis, ``Quantum supremacy using a programmable superconducting processor,'' \emph{Nature}, vol. 574, p. 505–510, 2019. [Online].
  Available: \url{https://www.nature.com/articles/s41586-019-1666-5}
\BIBentrySTDinterwordspacing

\bibitem{Pearson2020Simulating}
N.~J. Pearson, ``Simulating many-body quantum systems: Quantum algorithms and experimental realisation,'' Ph.D. dissertation, ETH Zurich, 2020.

\bibitem{Hirano2024MagicPool}
\BIBentryALTinterwordspacing
Y.~Hirano, Y.~Suzuki, and K.~Fujii, ``Magicpool: Dealing with magic state distillation failures on large-scale fault-tolerant quantum computer,'' 2024. [Online]. Available: \url{https://arxiv.org/abs/2407.07394}
\BIBentrySTDinterwordspacing

\bibitem{kobori2024lsqca}
T.~Kobori, Y.~Suzuki, Y.~Ueno, T.~Tanimoto, S.~Todo, and Y.~Tokunaga, ``Lsqca: Resource-efficient load/store architecture for limited-scale fault-tolerant quantum computing,'' \emph{arXiv preprint arXiv:2412.20486}, 2024.

\bibitem{Gidney2021howtofactorbit}
\BIBentryALTinterwordspacing
C.~Gidney and M.~Eker{\aa{}}, ``How to factor 2048 bit {RSA} integers in 8 hours using 20 million noisy qubits,'' \emph{{Quantum}}, vol.~5, p. 433, Apr. 2021. [Online]. Available: \url{https://doi.org/10.22331/q-2021-04-15-433}
\BIBentrySTDinterwordspacing

\bibitem{holmes2019resource}
A.~Holmes, Y.~Ding, A.~Javadi-Abhari, D.~Franklin, M.~Martonosi, and F.~T. Chong, ``Resource optimized quantum architectures for surface code implementations of magic-state distillation,'' \emph{Microprocessors and Microsystems}, vol.~67, pp. 56--70, 2019.

\bibitem{akahoshi2024compilation}
Y.~Akahoshi, R.~Toshio, J.~Fujisaki, H.~Oshima, S.~Sato, and K.~Fujii, ``Compilation of trotter-based time evolution for partially fault-tolerant quantum computing architecture,'' \emph{arXiv preprint arXiv:2408.14929}, 2024.

\bibitem{gidney2024inplace}
C.~Gidney, ``Inplace access to the surface code y basis,'' \emph{Quantum}, vol.~8, p. 1310, 2024.

\bibitem{Hirano2025LAPBCSimulation}
\BIBentryALTinterwordspacing
Y.~Hirano, ``Locality-aware pauli-based computation simulation source code.'' [Online]. Available: \url{https://github.com/yutakahirano/lapbc}
\BIBentrySTDinterwordspacing

\bibitem{leblond2025quantum}
T.~LeBlond and R.~S. Bennink, ``Quantum resource comparison for two leading surface code lattice surgery approaches,'' \emph{arXiv preprint arXiv:2506.08182}, 2025.

\end{thebibliography}

\end{document}